\documentstyle[12pt,epsf]{article}
\newcommand{\dif}[2]{\frac{d #1}{d #2}}
\def\x{{\left(\frac{\Lambda}{\mu}\right)}}

\begin{document}
\begin{titlepage}
  \begin{flushright}
    KUNS-1635\\[-1mm]
    hep-ph/0002102
  \end{flushright}
  \begin{center}
    \vspace*{1.4cm}
    
  {\Large\bf Yukawa hierarchy from extra dimensions and infrared fixed
    points} 
  \vspace{1cm}
  
  {Masako Bando\footnote{E-mail address: bando@aichi-u.ac.jp},
    Tatsuo Kobayashi\footnote{E-mail address: 
      kobayash@gauge.scphys.kyoto-u.ac.jp}, 
    Tatsuya Noguchi\footnote{E-mail address:
      noguchi@gauge.scphys.kyoto-u.ac.jp}, and
    Koichi Yoshioka\footnote{E-mail address:
      yoshioka@gauge.scphys.kyoto-u.ac.jp}}
  \vspace{5mm}
  
  $^*$ {\it Aichi University, Aichi 470-0296, Japan}\\
  $^{\dagger,\ddagger,\S}$ {\it Department of Physics, Kyoto
    University Kyoto 606-8502, Japan}
  \vspace{1.5cm}
  
  \begin{abstract}
    We discuss the existence of hierarchy of Yukawa couplings in the
    models with extra spatial dimensions. The hierarchical structure
    is induced by the power behavior of the cutoff dependence of the
    evolution equations which yield large suppressions of couplings at
    the compactification scale. The values of coupling constants at
    this scale can be made stable almost independently of the initial
    input parameters by utilizing the infrared fixed point. We find
    that the Yukawa couplings converge to the fixed points very
    quickly because of the enhanced energy dependence of the
    suppression factor from extra dimensions as well as in the case of
    large gauge couplings at high-energy scale.
  \end{abstract}
\end{center}
\end{titlepage}
\setcounter{footnote}{0}

It has long been one of the challenging subjects why the Nature
provides very small ratios of physical parameters. A typical example
is seen in the Yukawa-Higgs couplings which are small compared with
the strong gauge coupling, except for the top quark, and show quite
apparently a hierarchical structure between generations. There are
also many examples of coupling terms which should be hierarchically
suppressed, for example, neutrino masses, $R$-parity violating
interactions, and so forth. The perturbative analysis shows that the
coupling constants vary as functions of energy scale by radiative
corrections. However, in ordinary four-dimensional theories, there
emerges no large hierarchy because the corrections are only
logarithmically dependent on the energy scale. Therefore the observed
hierarchies must be given by the initial conditions at the high-energy
scale. 

Recently various phenomenological problems have intensively been
studied in the models with extra dimensions beyond the usual four
dimensions \cite{ED}--\cite{exp}. In these attempts, the coupling
constants show power-law running behavior by the contributions from
Kaluza-Klein modes if gauge and matter fields live in the extra
dimensions \cite{tv}--\cite{kkmz}. Using this power-law behavior, we
could obtain the hierarchical structures of couplings observed in the
low-energy region. It seems, however, that the low-energy coupling
constants largely depend on the initial conditions at the cutoff scale
and the predictability of models is lost due to the steep changes with
the energy scale.

In this letter, we discuss the existence of hierarchical Yukawa
couplings stabilized by infrared fixed points within the framework of
theories with extra dimensions. We also find the possibility of
applying the fixed point approach even in asymptotically free gauge 
theories, though in four dimensions such a situation usually occurs
only in asymptotically non-free theories.

Since we are interested in the possibility of large Yukawa hierarchy
in the extra dimension scenarios, we work within a generic setup and
investigate its infrared fixed point structure. Our setup covers the
situations that the matter and gauge fields live even on the branes 
spreading in different number of dimensions. Once the field
configuration is specified, we can write down the renormalization
group equations of the system and calculate the low-energy values of
running coupling constants. In this letter, we assume the generic form
of renormalization group equations, not referring to a specific
model. The model-dependent details and the applications to the mass
hierarchy between generations will be discussed in \cite{bkny}.

Before demonstrating the evolution of couplings in the extra
dimensions, let us briefly review the renormalization group equations
in a simple four-dimensional supersymmetric model containing
superfields $\Psi_1$, $\Psi_2$ and $\Psi_3$ which can couple to a
gauge multiplet with the coupling $g$, and we assume the following
superpotential
\begin{eqnarray}
  W &=& y\Psi_1\Psi_2\Psi_3.
\end{eqnarray}
The renormalization group equations of this system with a cutoff 
$\Lambda$ can be written as
\begin{eqnarray}
  \dif{\alpha}{t} &=& -\frac{b}{2\pi}\alpha^2, \label{4dgauge}\\
  \dif{\alpha_y}{t} &=& \frac{\alpha_y}{2\pi}(c\alpha-a\alpha_y),
  \label{4dyukawa} 
\end{eqnarray}
where
\begin{equation}
  \alpha \equiv \frac{g^2}{4\pi},\quad \alpha_y\equiv\frac{y^2}{4\pi},
  \quad t = \ln\frac{\Lambda}{\mu}. 
\end{equation}
Note that the coefficients $b$ and $c$ $(\ge 0)$ are constants of
order one expressed by the group indices, and $a$ $(>0)$ denotes the
multiplicity of the Yukawa coupling $y$ in the anomalous
dimensions. The system is asymptotically free (asymptotically
non-free) for $b<0$ ($b>0$). The analytic solutions of these
differential equations are obtained, 
\begin{eqnarray}
  \alpha(t) &=& \frac{\alpha(0)}{\displaystyle{1+\frac{b}{2\pi}
      \alpha(0)t}},\label{4dgaugeS}\\ 
  \alpha_y(t) &=& \frac{\alpha_y(0)E(t)}{1
    +\displaystyle{\frac{a}{2\pi}\alpha_y(0)F(t)}}\label{4dyukawaS}, 
\end{eqnarray}
where the functions $E(t)$ and $F(t)$ are defined as 
\begin{eqnarray}
  E(t) &=& \left(1+\frac{b}{2\pi}\alpha(0) t\right)^{c/b}
  = \left(\frac{\alpha(0)}{\alpha(t)}\right)^{c/b}, \label{E}\\
  F(t) &=& \int^t_0 E(t')dt'.
\end{eqnarray}
In order to analyze a fixed point structure, now we define the ratio 
of couplings $R$. From Eqs.\ (\ref{4dgauge}) and (\ref{4dyukawa}), we
have
\begin{eqnarray}
  \dif{R}{t} &=& \frac{-1}{2\pi}(b+c)\alpha R(R - 1),
  \label{diffR}\\[1mm]
  R &\equiv& \frac{a}{b+c}\frac{\alpha_y}{\alpha}.
\end{eqnarray}
That implies that $R^*=1$ is the infrared fixed point. Solving 
Eq.\ (\ref{diffR}), we obtain
\begin{eqnarray}
  \frac{R(t)-1}{R(t)} &=& \xi\,\frac{R(0)-1}{R(0)}, \label{R}
\end{eqnarray}
where $\xi$ is defined as
\begin{eqnarray}
  \xi &=& \frac{1}{E(t)}\left(\frac{\alpha(t)}{\alpha(0)}\right).
  \label{xi}
\end{eqnarray}
The suppression factor $\xi$ is written only by the gauge coupling
constant and provides the rate at which $R$ approaches to the infrared
fixed point value $R^*=1$, which has been discussed to demonstrate the
fast convergency in the asymptotically non-free gauge 
models \cite{anf}. Using Eq.\ (\ref{E}), we 
have $\xi=\left(\alpha(t)/\alpha(0)\right)^{1+c/b}$ which shows that
when $b>0$, $\xi\to 0$ very rapidly with increasing $t$ (in the
infrared). On the other hand, $\xi$ is not so suppressed in the case
of asymptotically free theories ($b<0$) and then the predictions from
the fixed point approach cannot be reliable.

Now let us consider a gauge-Yukawa system in higher dimensional
spacetime assuming that the chiral superfields as well as the gauge
multiplet can live in $4+\delta$ dimensions with $\delta$ dependent on
each field. The number of dimensions of the extended space may be more
than one and it becomes even 6 in string theories. Thus each
compactification scale may be different from the others, but we here
suppose that for simplicity, all the compactification scales are 
equal to $\mu_0$ although the following analyses can be applied to
generic cases straightforwardly. Above the compactification 
scale $\mu_0$, the contributions of the Kaluza-Klein modes come into 
play. Neglecting the logarithmic terms from the contributions of
ordinary four-dimensional particles, the truncated 1-loop effective
renormalization group equations are written as \cite{power1}
\begin{eqnarray}
  \dif{\alpha}{t} &=& -\frac{\widetilde b}{2\pi} X_{\delta_g}
  \x^{\delta_g}\alpha^2 +\cdots, \label{EDgauge}\\
  \dif{\alpha_y}{t} &=& \frac{\alpha_y}{2\pi}\sum_{i=1,2,3}
  \gamma_i, \label{EDyukawa} 
\end{eqnarray}
where $\gamma_i$ denote the anomalous dimensions of $\Psi_i$,
\begin{eqnarray}
  \gamma_i &=& \widetilde c_i X_{\delta_{g_i}} \x^{\delta_{g_i}}\alpha
  -\widetilde a_i X_{\delta_i} \x^{\delta_i}\alpha_y +\cdots
\end{eqnarray}
with the cutoff $\Lambda$, above which effects of new physics become
important. In the above equations, the power factors appeared in the
beta functions originate from the Kaluza-Klein modes propagating in
the loops. The $\delta_g$ part corresponds to the largest gauge
contribution to the anomalous dimension of gauge fields and the
ellipses denote less dominant terms with smaller powers, which have
negligible effects on the evolution of couplings. Similarly,
$\delta_{g_i}$ and $\delta_i$ are the largest gauge and Yukawa
contributions, respectively, to the anomalous dimension of matter
fields. These exponents can be determined once we fix the
configuration of relevant fields in the extra dimensions. The volume
factor $X_{\delta}=\pi^{\delta/2}/\Gamma(1+\delta/2)$ is originated
from the phase-space integral of Kaluza-Klein modes.\footnote{In 
  Ref.\ \cite{kkmz}, it has been 
  obtained $X_\delta=\pi^{\delta/2}/\Gamma(2+\delta/2)$. The 
  difference can be absorbed by the redefinition of the cutoff
  $\Lambda$ for $\delta_i=\delta_g$. Our results are irrelevant to the
  explicit value of $X_\delta$.} 
In the following, we redefine the coefficients in the beta functions
for simplicity as $a_i=\widetilde a_i X_{\delta_i}$, 
$b=\widetilde b X_{\delta_g}$, and $c_i=\widetilde c_i
X_{\delta_{g_i}}$.

Our goal is to show how the hierarchy of Yukawa couplings is realized
as a result of their running effects from $\Lambda$ to $\mu_0$. Thus
$\mu_0$ is regarded as the energy scale at which we expect that Yukawa
couplings are determined almost independently of the initial values 
at $\Lambda$. This may be possibly realized if they have infrared
fixed points. For that purpose, we now investigate the fixed point
structure at the Kaluza-Klein threshold scale $\mu_0$. From 
Eqs.\ (\ref{EDgauge}) and (\ref{EDyukawa}), we have 
\begin{eqnarray}
  \dif{}{t}\ln\frac{\alpha_y}{\alpha} &=& -\frac{1}{2\pi}
  \left[a \x^{\delta_y}\alpha_y -\left\{b\x^{\delta_g}
      +c\x^{\delta_{g'}}\right\}\alpha\right]
\end{eqnarray}
with $\delta_y= {\rm Max}(\delta_i)$ and 
$\delta_{g'}= {\rm Max}(\delta_{g_i})$, and $a$ and $c$ are the
corresponding coefficients. By the analogy with the four-dimensional
case, we define 
\begin{eqnarray}
  R &\equiv& \frac{a (\Lambda/\mu)^{\delta_y}}{b
    (\Lambda/\mu)^{\delta_g} +c (\Lambda/\mu)^{\delta_{g'}}} 
  \cdot\frac{\alpha_y}{\alpha},
\end{eqnarray}
then the renormalization group equation for $R$ becomes
\begin{eqnarray}
  \dif{R}{t} &\simeq& -\frac{1}{2\pi}\left[b\x^{\delta_g} 
    +c \x^{\delta_{g'}} \right]\alpha R(R-1).
\end{eqnarray}
This equation shows that $R^*=1$ is the infrared stable fixed point of
this model as long as $R$ is positive. It should be noted that 
while $R^*$ is the constant, the `fixed point' value 
of $\alpha_y/\alpha$ is generally energy dependent. In the above
equation, we have neglected the sub-leading terms suppressed by the
small value of $\mu/\Lambda$. If one includes these terms, the fixed
point value $R^*=1$ receives a small correction suppressed 
by $\mu/\Lambda$ but then the convergence behavior to that fixed point
by the factor $\xi$ is unchanged. In other words, even when one
redefines $R$ such that $R\to 1$ in the infrared, this change has
little effect on the suppression factor $\xi$.

We can find how fast $R$ approaches to the infrared fixed point by
solving the above renormalization group equation. The solutions 
of $\alpha_y$, $R$ and the suppression factor $\xi$ obtained in the
four-dimensional case ((\ref{4dyukawaS}), (\ref{R}) and (\ref{xi}))
still hold in the extra dimension scenarios if we now replace the
definitions of $E(t)$ and $F(t)$ with
\begin{eqnarray}
  E(t) &=& \exp\left[\int^t_0 dt'\,\frac{c}{2\pi}
    \x^{\delta_{g'}}\alpha\right],\\
  F(t) &=& \int^t_0 dt'\, \x^{\delta_y} E(t').
\end{eqnarray}
It is easily seen that in the case of $\delta_y=\delta_{g'}=\delta_g$,
the forms of the solutions are identical to those in four dimensions
with a suitable change of the variable $t$.

\begin{figure}[htbp]
  \begin{center}
    \leavevmode
    \epsfxsize=11cm \ \epsfbox{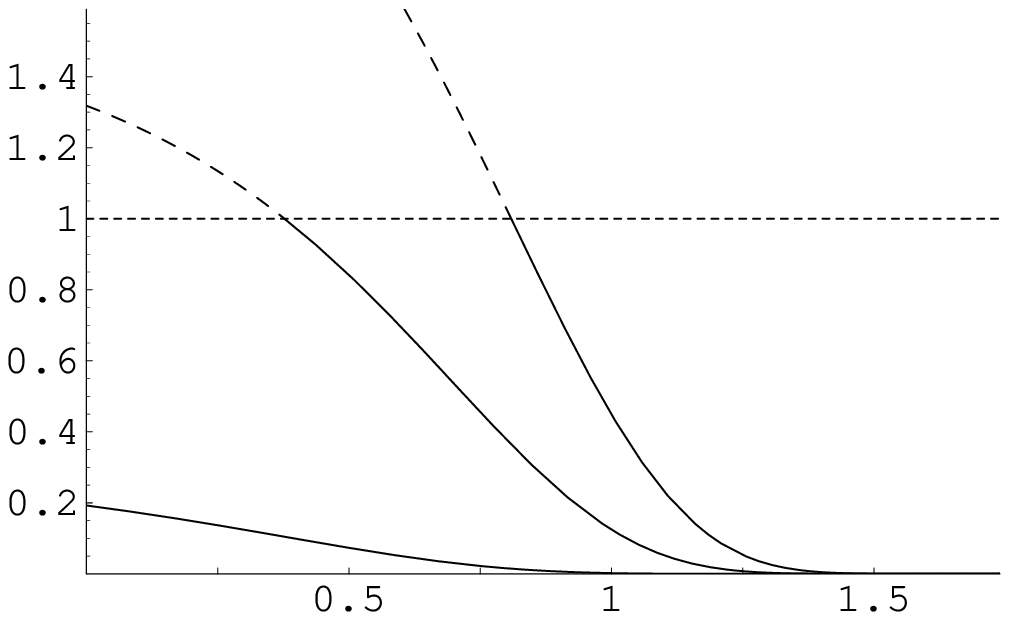}
    \put(7,20){\large $\delta_{g'}/\delta_g$}
    \put(-289,208){$\xi$}
    \put(-186,112){$b=-3$}
    \put(-212,76){$b=-1$}
    \put(-240,42){$b=1$}
    \caption{The typical behaviors of the suppression factor
      $\xi$. Above the horizontal dotted line $\xi=1$, the fixed-point
      solution becomes infrared unstable. $\delta_{g'}/\delta_g=1$
      recovers the four-dimensional values of $\xi$. We set
      $\alpha(t)=0.1$ and $c=16/3$.}
    \label{suppression}
    \vspace*{15mm}

    \epsfxsize=11cm \ \epsfbox{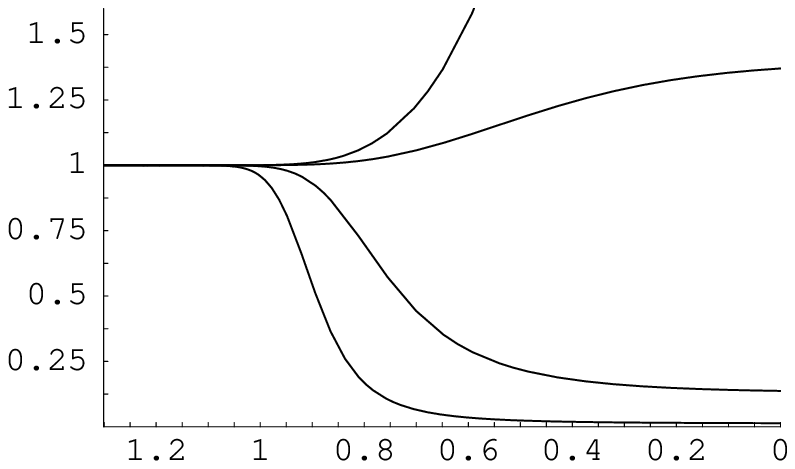}
    \put(5,12){$\ln\,(\Lambda/\mu)$}
    \put(-275,197){$R$}
    \put(-121,190){$\alpha_y(0)=1$}
    \put(2,163){$\alpha_y(0)=0.1$}
    \put(2,42){$\alpha_y(0)=0.01$}
    \put(2,27){$\alpha_y(0)=0.001$}
    \caption{The renormalization group evolutions of $R$ approaching
      to the fixed point $R^*=1$ with various initial conditions of
      $\alpha_y$. We set $\delta_g=1$, $\delta_{g'}=\delta_y=2$,
      $a=3$, $b=-3$, and $c=16/3$.}
    \label{rt}
  \end{center}
\end{figure}

Since the couplings $\alpha$ and $\alpha_y$ have power running
behaviors, the running region between $\Lambda$ and $\mu_0$ may be
much narrower than the four-dimensional case. Therefore if we make use
of the infrared fixed points, the strong convergence to them is
required. We now examine the behavior of the suppression 
factor $\xi=E(t)^{-1}(\alpha(t)/\alpha(0))$ when the power 
factors $\delta_y$, $\delta_g$ and $\delta_{g'}$ are varied. First, it
is noted that $\xi$ is determined only from the informations of the
gauge sector, $\delta_{g,g'}$ and $\alpha$, but does not depend 
on $\delta_y$ \cite{fermion}. This means that the convergency to the
infrared fixed point is dominated by the gauge coupling constant
(except for the large $\alpha_y(0)$ case). In Fig.\ \ref{suppression},
we show the dependence of the suppression factor $\xi$ on
$\delta_{g'}/\delta_g$ and $b$. From this figure, one can see that
there are two distinct ways to have a small value of $\xi$. One is
realized when $b>0$, namely in the asymptotically non-free
theories. This is clearly seen when the effect from the gauge
anomalous dimension is dominant in the evolution 
of $R$ ($\delta_{g'}/\delta_g <1$). The situation is almost similar to
the ordinary four-dimensional cases \cite{anf}. The other possibility
is essentially due to the existence of the extra dimensions. When the
gauge contributions of the matter anomalous dimensions govern the
renormalization group equation ($\delta_{g'}/\delta_g>1$), the
suppression factor $\xi$ becomes very small even in the asymptotically
free theories such as the minimal supersymmetric standard model. It
means that $R$ approaches its fixed-point value very quickly in the
infrared. This is in sharp contrast to the four-dimensional
case. Figure \ref{rt} shows that $R$ indeed converges to the infrared
fixed point ($R^*=1$) within the region $\Lambda/\mu<O(10)$ for a wide
range of initial Yukawa couplings. If $\xi$ approaches to 0 rapidly
enough, on the infrared point $\mu_0$ we have a relation between
$\alpha_y(\mu_0)$ and $\alpha(\mu_0)$ independently of the high-energy
input values.
\begin{eqnarray}
  \alpha_y^*(\mu_0) &=& \left[\frac{b}{a}
    \left(\frac{\mu_0}{\Lambda}\right)^{\delta_y-\delta_g}
    +\frac{c}{a} 
    \left(\frac{\mu_0}{\Lambda}\right)^{\delta_y-\delta_{g'}}
  \right]\alpha^*(\mu_0).
  \label{fp}
\end{eqnarray}
Thus we can obtain a variety of hierarchical power factors 
for $\mu_0/\Lambda$ in the infrared fixed points of Yukawa couplings
according to the values of $\delta$'s.

When $\xi$ is not so small, by contrast, the system flows slowly to
the fixed point and then Eq.\ (\ref{R}) shows that the initial $R(0)$
dependence of $R(t)$ generally spoils the predictions from the
infrared fixed points. However, even in this case, if $R(0)$ is large
enough, the $R(0)$ dependence on the right-handed side of 
Eq.\ (\ref{R}) disappears and the predictability is recovered. In this
case, we have
\begin{eqnarray}
  R^*(\mu_0) &=& \frac{1}{1-\xi(\mu_0)}.
\end{eqnarray}
This corresponds to the so-called quasi fixed point \cite{qfp}. As
well as in the true fixed point approach, we can still obtain a
hierarchical structure of Yukawa couplings from the fixed points,
i.e., independently of the high-energy input parameters.

\begin{figure}[htbp]
  \begin{center}
    \leavevmode
    \epsfxsize=11cm \ \epsfbox{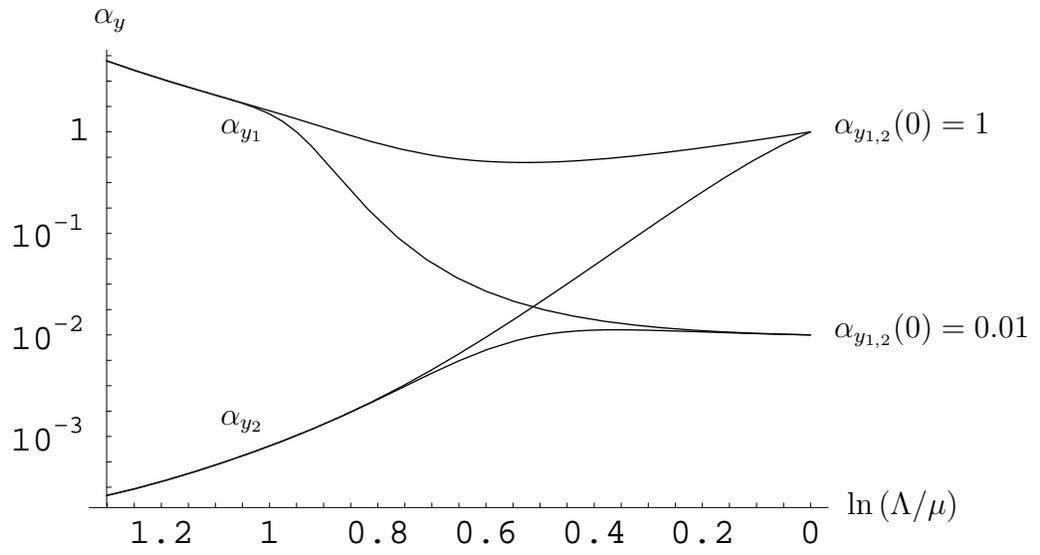}
    \put(7,16){$\ln\,(\Lambda/\mu)$}
    \put(-278,202){$\alpha_y$}
    \put(2,160){$\alpha_{y_{1,2}}(0)=1$}
    \put(2,83){$\alpha_{y_{1,2}}(0)=0.01$}
    \put(-230,160){$\alpha_{y_1}$}
    \put(-230,50){$\alpha_{y_2}$}
    \caption{The fixed point hierarchy of Yukawa couplings in the two
      generation model. The initial values of two couplings at
      $\mu=\Lambda$ are taken to be equal, $\alpha_{y_{1,2}}(0)=1$ or
      0.01. We set $\delta_{y_1}=1$ and $\delta_{y_2}=4$ and the
      other beta functions are same as in Fig.\ \ref{rt}.}
  \label{2yukawa}
\end{center}
\end{figure}

So far, we have considered the simple model with only one Yukawa
coupling. We will extend the model to contain more numbers of Yukawa
couplings and show that a large hierarchy is indeed generated.
For simplicity, consider a case with two independent Yukawa
couplings $\alpha_{y_1}$ and $\alpha_{y_2}$. As seen from 
Eq.\ (\ref{fp}), the hierarchy among them comes from different choices
of $\delta$'s. In Fig.\ \ref{2yukawa}, we show an example of possible
hierarchy in this two generation model 
with $\delta_g=1$, $\delta_{g'_1,g'_2}=2$, $\delta_{y_1}=1$, 
and $\delta_{y_2}=4$. Since we take $\delta_{g'_i}>\delta_g$, the
infrared fixed points are actually realized even in $b<0$
(asymptotically free case). The model gives rise to a large 
hierarchy $\alpha_{y_1}/\alpha_{y_2}\sim10^4$ at the Kaluza-Klein
threshold $\mu_0$ due to the difference of $\delta_y$. As discussed
above, these values correspond to the fixed points and not dependent
on the initial values $\alpha_{y_i}(0)$. It is straightforward to
extend the above discussions to the models with more numbers of Yukawa
couplings. This interesting possibility may be readily applied to the
quarks and leptons mass hierarchy between generations and to other
phenomenological problems involving very different order of coupling
constants \cite{bkny}.

Here a few comments on the available value of $\Lambda/\mu_0$ are in
order. In Ref.\ \cite{fermion}, the relation between the order of
hierarchy factor and the perturbation limit was emphasized. They
concluded that if one uses the infrared fixed points to have
hierarchical Yukawa couplings, one is forced to adopt the strong
unification scenarios and then the perturbative reliability restricts
the hierarchy factor to less than the order of 10. However, we have
seen in the above that the strong convergency to the Yukawa fixed
points can be accomplished not only in the case of the strong gauge
couplings but also $\delta_{g'}/\delta_g>1$. The latter is realized of
course even in asymptotically free models and thus, the perturbation
limit can be rather relaxed. Secondly, if one tries to incorporate the
fixed point scenario with the gauge coupling unification, it seems
natural to identify $\Lambda$ to the unification scale. In that case,
the largest value of available hierarchy is roughly estimated as
\begin{eqnarray}
  \left(\frac{\Lambda}{\mu_0}\right)^{\delta_g} &\sim& 
  \ln\left(\frac{M_{\rm GUT}}{M_W}\right) \;\sim\; 30,
\end{eqnarray}
which also constrains the value of $\Lambda/\mu_0$, and we may require
a small $\delta_g$ in order to obtain an enough large $\Lambda/\mu_0$ 
for the quarks and leptons mass differences. After all, to satisfy the
above two limits one needs a smaller value of $\delta_g$ compared to
other power factors. This could be achieved by some cancellations
between diagrams, the specific field configurations in the extra
dimensions, and so on. It should be noted that the asymmetrical
compactifications of extra dimensions can also lead a various orders
of hierarchy factors from infrared fixed points even with the common
values of $\delta$'s.

We have studied the infrared fixed points behaviors of the
renormalization group equations of a simple gauge-Yukawa model with
extra dimensions. It is found that the hierarchies among the Yukawa
couplings can be realized as the fixed point values. The fixed-point
structures largely depend on the field configurations in the extra
dimensions ($\delta_i$, $\delta_g$ and $\delta_{g_i}$). Under certain
conditions concerning the extra dimensions, one can observe the strong
attraction to the infrared fixed points and then it can be applicable
to various models in which hierarchical structure of couplings is
needed. Though there might exist some subtleties in applying it to
phenomenological models, we hope this approach will give a possible
solution to unsolved problems in particle physics.

\subsection*{Acknowledgements}

The authors would like to thank T.\ Kugo and M.\ Nojiri for valuable
discussions and comments. M.\ B.\ and K.\ Y.\  are supported in part
by the Grants-in-Aid for Scientific Research No.\ 09640375 and the
Grant-in-Aid No.\ 9161, respectively, from the Ministry of Education,
Science, Sports and Culture, Japan.

\newpage

\end{document}